\documentclass[a4paper,11pt]{article}
\usepackage{jheppub} % use jheppub.sty

\usepackage{amsfonts,mathrsfs,textcomp,dsfont,esint,array,multirow,url,graphicx}
\usepackage[mathscr]{euscript}
\usepackage{tikz,tikz-cd}

\newcommand{\iso}{\cong}

\newcommand{\N}{\mathcal{N}}

\DeclareMathOperator{\tr}{tr}

\DeclareMathOperator{\diag}{diag}

\title{$\mathcal{N}=1$ Curves on Generalized Coulomb Branches}

\author[a]{Thomas Bourton,}
\author[a]{Elli Pomoni,}
\author[a]{Xinyu Zhang}
\affiliation[a]{Deutsches Elektronen-Synchrotron DESY, Notkestr. 85, 22607 Hamburg, Germany}

\preprint{DESY 21-173}

\abstract{
We study the low energy effective dynamics of four-dimensional  $\mathcal{N}=1$ supersymmetric gauge theories of class  $\mathcal{S}_k$  on the generalized Coulomb branch.
The low energy effective gauge couplings are naturally encoded in algebraic curves $\mathcal{X}$, which we derive for general values of the couplings and mass deformations.
We then recast these IR curves $\mathcal{X}$ to the UV or M-theory form $\mathcal{C}$: the punctured Riemann surfaces on which the six-dimensional $\mathcal{N}=(1,0)$ ${A_{k-1}}$ SCFTs are compactified giving the class $\mathcal{S}_k$ theories. 
We find that the UV curves $\mathcal{C}$ and their corresponding meromorphic differentials take the same form as those for their mother four-dimensional  $\mathcal{N}=2$ theories of class  $\mathcal{S}$. They have the same poles, and their residues are functions of all the exactly marginal couplings and the bare mass parameters which we can compute exactly.
}

\begin{document}
\maketitle

\section{Introduction}

Famously, the low energy dynamics of four-dimensional (4D) $\mathcal{N}=2$
supersymmetric gauge theories can be determined
using
the constraints of holomorphy, global symmetries
%, monodromies around the singularities, 
and consistency in various limits. The low energy effective action on the Coulomb branch is fully 
 encoded in the Seiberg-Witten curve  \cite{Seiberg:1994rs,Seiberg:1994aj}.
Shortly after the seminal work by Seiberg and Witten, Intriligator and Seiberg pointed out that similar techniques can be employed to study the infrared (IR) dynamics of $\N=1$ gauge theories as long as they possess an abelian Coulomb branch \cite{Intriligator:1994sm}.
In \cite{Tachikawa:2011ea,Maruyoshi:2013hja,Xie:2013rsa,Bonelli:2013pva,Giacomelli:2014rna,Xie:2014yya,Tachikawa:2018sae}  a modern approach to the $\N=1$ curves was taken. For $\N=1$ theories engineered by wrapping M5 branes on a Riemann surface which is embedded in a local Calabi-Yau threefold \cite{Bah:2012dg}, their IR curves are given by the spectral 
curves of the generalized Hitchin systems
which involve a pair of commuting Hitchin fields.

$\N=1$ SCFTs of class $\mathcal{S}_k$  \cite{Gaiotto:2015usa} can be obtained via orbifolding  $\N=2$ SCFTs of Class $\mathcal{S}$  \cite{Gaiotto:2009we,Gaiotto:2009hg}. They can also be engineered using M-theory on
$\mathbb{R}^{3,1}\times (CY_2 \times \mathbb{C})/\mathbb{Z}_k \times \mathbb{R}$ 
with M5 branes lying along $\mathbb{R}^{3,1}\times \mathcal{C}$ with the Riemann surface $\mathcal{C} \subset CY_2 /\mathbb{Z}_k $.
It is an active field of research to examine $\N=1$ theories of class $\mathcal{S}_k$ and the broader theories of class $\mathcal{S}_\Gamma$ where the $\mathbb{Z}_k$ singularity is generalized to any ADE orbifold singularity $\Gamma$ \cite{Coman:2015bqq,Mitev:2017jqj,Bourton:2017pee,Bah:2017gph,Razamat:2018zus,Bourton:2020rfo,Razamat:2016dpl,Heckman:2016xdl,Kim:2017toz,Kim:2018lfo,Kim:2018bpg,Apruzzi:2018oge,Razamat:2019ukg,Razamat:2020bix,Chen:2021ivd,Nazzal:2021tiu}.

A supersymmetric theory
often possesses several inequivalent supersymmetry-preserving ground states
which give rise to the so called moduli space of supersymmetric vacua of the theory. In many cases the moduli space is a manifold (possibly with singularities) parametrized by the vacuum expectation values (vevs) of a set of gauge-invariant operators.
Depending on the behavior of the potential for test charges, the moduli space can generically be classified into different `branches', leading to different `phases' of the theory. These phases may be Coulomb, Higgs, confining or, more generally, a mixture between them. 

$\N=2$ theories, due to their large R-symmetry, possess distinct Coulomb and Higgs branches (as well as mixed branches).
For generic  $\N=1$ theories it is not possible to separate distinct branches
of supersymmetric vacua,
thus the study of their moduli space is in general complicated.
Nonetheless, it has been recently understood \cite{Coman:2015bqq,Mitev:2017jqj,Bourton:2017pee,Razamat:2018zus,Bourton:2020rfo} that there is a special but very broad and interesing class of $\N=1$ superconformal theories (SCFTs),
the so called $\N=1$ theories of class $\mathcal{S}_k$ \cite{Gaiotto:2015usa}, for which it is possible to distinguish between generalized Coulomb and Higgs branches precisely as for $\N=2$ gauge theories. 
This is due to the fact that in addition to the R-symmetry they enjoy, by construction, extra global symmetries. Consequently, a generalized Coulomb branch can be isolated and has been successfully studied \cite{Coman:2015bqq,Mitev:2017jqj,Bourton:2017pee,Razamat:2018zus,Bourton:2020rfo}.

For theories of class $\mathcal{S}_k$, the $\N=1$ curves 
 on the generalized Coulomb branch simplify and instead of a generalized Hitchin system, they are associated to a usual  Hitchin system with only one  Hitchin field  \cite{Coman:2015bqq}.
 This is because as for  $\N=2$ theories we can isolate the  $CY_2/\mathbb{Z}_k$ piece of the geometry to be $T^* \mathcal{C}$.
 What is more,  in \cite{Coman:2015bqq} 
 the class $\mathcal{S}_k$ $\N=1$ curves 
 were derived at special points of the conformal manifold,  the so called orbifold points, where the coupling constants arising from the same gauge group of the mother theory in class $\mathcal{S}$ are taken to be equal.
One of our goals in this paper is to obtain these curves at a generic point of the conformal manifold using the techniques developed in \cite{Intriligator:1994sm}.

The construction of \cite{Intriligator:1994sm}  produces the curves in a form which can be thought of as
 the IR form, capturing low energy data on the  generalized Coulomb branch. 
 Their derivation  is presented   in Section \ref{sec:SI4classSk}.
In this paper we will denote the IR curves by $\mathcal{X}$.
Following the seminal work of Gaiotto  \cite{Gaiotto:2009we} we can bring an IR curve $\mathcal{X}$ to the Gaiotto form, which is the UV or M-theory curve $\mathcal{C}$.  
The UV curve $\mathcal{C}$ describes the Riemann surface on which $N$ M5 branes wrap in the M-theory realization of theories of class $\mathcal{S}_k$, and is a ramified cover of the IR curve $\mathcal{X}$  \cite{Gaiotto:2015usa,Coman:2015bqq}.
Doing so, in Section \ref{sec:UVcurves},
we find that these curves and the associated meromorphic differentials have a very similar pole structure
 as their mother
 $\mathcal{N}=2$ theories of class  $\mathcal{S}$. The novel (genuinely $\mathcal{N}=1$) information is contained in 
the residues of the meromorphic differentials. They capture the mass deformations, and turn out to be functions of all the exactly marginal couplings and the bare mass parameters. This can be interpreted as a finite renormalization of the bare masses that appear in the Lagrangian.
Amazingly, we are able to compute these renormalized masses exactly.

%%%%%%%%%%%%%%%%%%%%%%%%%%%%%%%%%%%%%%%%%%

\section{\texorpdfstring{$\mathcal{N}=1$}{N=1} Intriligator-Seiberg Curves}
In this section we wish to  briefly review  the techiques of   Intriligator and Seiberg \cite{Intriligator:1994sm} together with
a few examples of curves which we will use in Section \ref{sec:SI4classSk}.

\subsection{General principles}

At a generic point of the generalized Coulomb branch $\mathbf{CB}$ of the moduli space, the low energy effective
theory is described in terms of $r$ $\mathcal{N}=1$ abelian vector
multiplets with the associated field strength superfields $W_{\alpha}^{a},a=1,\cdots,r$ (where $r$ is the rank of the gauge group),
and possibly neutral moduli fields $U_{I}$. In terms of $\N=1$ superspace, the gauge kinetic term in the low energy effective action takes the form
\begin{equation}
\mathcal{S}_{eff}=\frac{1}{16\pi}\mathrm{Im}\int d^{4}xd^{2}\theta\,\tau_{ab}\left(q_{i},U_{I}\right)W_{\alpha}^{a}W^{\alpha b}+\cdots,
\end{equation}
where $\tau=\left(\tau_{ab}\right)$ with ${a,b=1,\cdots,r}$ is the matrix
of the effective holomoprhic gauge couplings. Supersymmetry requires that $\tau$
must be holomorphic in all the holomorphic coupling constants $q_{i}$ of the
underlying microscopic theory and in the neutral chiral superfields $U_{I}$.
Since the massless photons are subject to the electric-magnetic duality,
any $\mathrm{Sp}\left(2r,\mathbb{Z}\right)$ transformation
of $\tau$ leaves physics invariant. On the other hand, $\tau$ is
not single-valued, and can have nontrivial monodromies as one changes
$q_{i}$ and $U_{I}$ along a closed path. Hence, $\tau$ should be
viewed as a section of an $\mathrm{Sp}\left(2r,\mathbb{Z}\right)$
bundle over the parameter space of $q_{i}$ and $U_{I}$. 

Following \cite{Seiberg:1994rs,Seiberg:1994aj,Intriligator:1994sm},
it is convenient to identify $\tau$  with a normalized period matrix
of an algebraic curve of genus $r$. This curve is called the IR $\mathcal{N}=1$
curve and in this paper we will denote it by $\mathcal{X}$. 
It is given by an equation in two complex variables $x,y$,
\begin{equation}
\mathcal{X}:\quad y^2=F(x;q_i,U_{I},m_f)\,,
\end{equation}
where $F(x;q_i,U_{I},m)$ is a polynomial in $x$ of degree $2r+1$ or $2r+2$, with the coefficients depending on $q_i$, $U_{I}$, and masses $m_f$. 
The matrix $\tau$ of the effective gauge couplings is then computed from
\begin{equation}
\tau_{ab} \oint_{\beta^b} \omega_c = \oint_{\alpha_a} \omega_c \,,
\end{equation}
where we have picked a symplectic homology basis $\left\{\alpha_1, \cdots, \alpha_r; \beta^1,\cdots,\beta^r\right\}$ for $\mathcal{X}$, and the canonical basis of holomorphic differentials $\omega_c = x^{c-1}/y$. 

The $\mathcal{N}=1$ curve $\mathcal{X}$ should be compatible with all the global symmetries of the
theory, and becomes singular at the submanifolds of the moduli space
where extra charged particles become massless. One can determine the electric
and magnetic charges of these charged massless particles from the
monodromies associated with the singularities. Only if all charged
massless particles are mutually local can the low energy effective
theory be a free field theory in an appropriate electric-magnetic duality frame.
Otherwise, the low energy effective theory will be an interacting
$\mathcal{N}=1$ superconformal field theory.

The curve $\mathcal{X}$ varies at different points of $\mathbf{CB}$. Therefore, it is useful to consider a family of holomorphic curves $\mathfrak{X}$, which is the total space of holomorphic curves fibered over the moduli space  $\mathbf{CB}$,
\begin{equation}
\mathfrak{X}=\begin{tikzcd}
\mathcal{X}\arrow[d]\\
\mathbf{CB}
\end{tikzcd}\,.
\end{equation}

There are major differences between the low energy solutions for genuine $\mathcal{N}=1$ theories
and those for $\mathcal{N}=2$ theories. First of all, one does not have a complete
solution to the low-energy dynamics for a genuine $\mathcal{N}=1$
theory, since the K\"ahler potential is not holomorphic and is therefore
not protected from quantum corrections. Second, the $\mathcal{N}=1$
curve may be not accompanied by a meromorphic differential which encodes
central charges and masses of BPS states.  To obtain the meromorphic differential, the string theory or M-theory description is usually needed \cite{Witten:1997ep,Maruyoshi:2013hja,Xie:2013rsa,Bonelli:2013pva,Giacomelli:2014rna,Xie:2014yya,Tachikawa:2018sae}.
If the $\mathcal{N}=1$ theory is engineered using M-theory on the geometry $\mathbb{R}^{3,1}\times CY_3  \times \mathbb{S}^1$,  we can further use the holomorphic three form 
 \begin{equation}
 \omega^{(3)} =   ds  \wedge dz \wedge dw 
 \end{equation}
where the map between the M-theory coordinates $s,z,w$ (with $s= - R_{M} \ln t$) and the IR coordinates $x,y$ will be specified in Section \ref{sec:UVcurves}.
The $CY_3$ is locally comprised of two line bundles fibered over the UV curve $\mathcal{L}_z \oplus \mathcal{L}_w \rightarrow \mathcal{C}$. For theories of class $\mathcal{S}_k$,  the first Chern class $c_1(\mathcal{L}_w) = 0$, allowing us to select an meromorphic two form $\omega^{(2)} \in\Omega^{2,0}(\mathcal{C})$,
\begin{equation} 
\omega^{(2)} = dz \wedge ds = d \omega \, .
\end{equation}
As described in \cite{Coman:2015bqq}, we can eventually obtain $\omega\in\Omega^{1,0}(\mathcal{X})$ after a parameter map which will be presented in Section \ref{sec:UVcurves}.

What is more, the moduli fields in $\mathcal{N}=1$ theories often satisfy intricate
relations, which are absent in $\mathcal{N}=2$ theories. Therefore,
one has to take one more step in determining the $\mathcal{N}=1$
curve, namely solving the relations among the moduli fields. 
Going to the generalized Coulomb branch, as we will do in this paper, one can avoid this complication.

\subsection{Examples}
\subsubsection{Pure $SU(2)$ $\mathcal{N}=2$ Gauge Theory}
The first example is the $\N=2$ supersymmetric Yang-Mills theory, which is the $\N=1$ supersymmetric gauge theory with a massless adjoint chiral multiplet. The curve that encodes the low-energy dynamics of the theory with gauge group $SU(2)$ is given by \cite{Seiberg:1994rs,Seiberg:1994aj}
\begin{equation}
y^2=x^3-ux^2+\frac{\Lambda^4}{4} x\,,
\end{equation}
where $u=\langle\mathrm{Tr}\phi^2\rangle$ is a gauge-invariant order parameter of the Coulomb moduli space, with $\phi$ the adjoint scalar field, and $\Lambda$ is the dynamically generated scale.
 
\subsubsection{$\mathcal{N}=1$ $SU(2)\times SU(2)$ Gauge Theory}
Perhaps the most relevant example for us 
that has been studied in \cite{Intriligator:1994sm} is the $\N=1$ gauge theory with gauge group $G=SU(2)_1\times SU(2)_2$ and two chiral superfields $\Phi_{ia_1a_2}$ in the $\left(\mathbf{2},\mathbf{2}\right)$ representation of $G$, labelled by $a_1,a_2=1,2$ respectively. Notice that the representation $\mathbf{2}$ of $SU(2)$ is pseudo-real and therefore isomorphic to the $\bar{\mathbf{2}}$. The theory has an $SU(2)_F$ flavour symmetry $\Phi_i\to f_{ij}\Phi_j$ where $f_{ij}\in SU(2)_F$. The theory can be understood as a $\mathbb{Z}_2$ orbifold of the above $\N=2$ pure $SU(2)$ gauge theory. Equivalently, the theory can be obtained by compactifying the 6d $(1,0)_{A_1}$ theory on a punctured Riemann surface $\mathcal{C}$ of genus zero in the presence of a certain set of defect operators associated to so called `wild' punctures. As in \cite{Gaiotto:2009we}, the $\mathcal{N}=1$ curve $\mathcal{X}$ is a double covering of $\mathcal{C}$. 

The theory has a three complex dimensional moduli space of supersymmetric vacua parametrized by the gauge invariant quantities
\begin{equation}
M_{ij}:=\det \Phi_i\Phi_j = \frac{1}{2}\Phi_{i a_1 a_2}\Phi_{j a_1^{\prime} a_2^{\prime}}\epsilon^{a_1 a_1^{\prime}}\epsilon^{a_2 a_2^{\prime}},
\end{equation}
where the determinant is taken over $G$.

The classical scalar potential vanishes along the D-flat directions,
\begin{equation}
\left\langle U^{-1} \begin{pmatrix} 0 & \Phi_1\\
\Phi_2 & 0
\end{pmatrix} U \right\rangle =\begin{pmatrix}v & 0\\
0 & -v 
\end{pmatrix} \otimes \begin{pmatrix}1 & 0\\
0 & -1 
\end{pmatrix},
\end{equation}
where $U$ is a $4\times 4$ unitary matrix, and $v$ is a complex number. Such field configurations do not break supersymmetry, but break the gauge group $G$ by the Higgs mechanism down to a subgroup $U(1)_D\subset SU(2)_D$, where $SU(2)_D$ is a diagonal embedding of $SU(2)_1\times SU(2)_2$. Therefore, the theory is in the abelian Coulomb phase with an one complex-dimensional moduli space. Due to the constraint of symmetries, no superpotential can be dynamically generated, and the classical vacuum degeneracy is not lifted quantum mechanically. Thus, the quantum theory also has an one complex-dimensional abelian Coulomb moduli space $\mathbf{CB}$ of vacua, parametrized by a single gauge and flavour singlet $u=\det_{ij}M_{ij}$.

The holomorphic effective gauge coupling is  $\tau=\tau\left(u,\Lambda_1^4,\Lambda_2^4\right)$, where $\Lambda_i^4:=\mu^4 e^{-8\pi^2/g_i^2(\mu)}$ are the dynamically generated scales. We may deduce the form of the curve describing the moduli space by considering two different limits. Firstly consider the limit where $\Phi_1$ acquires a large diagonal vev but the vev of $\Phi_2$ is vanishing. The gauge group is broken to $SU(2)_D$, and $\Phi_2$ decomposes into $\mathbf{2}\otimes\mathbf{2}\to \mathbf{3}\oplus\mathbf{1}$ of $SU(2)_D$. There is also a heavy singlet field $M_{11}$. If we decouple all the singlet fields, the theory is approximately pure $\N=2$ $SU(2)_D$ gauge theory, and the curve of that theory is
\begin{equation}\label{eqn:largeu}
y^2=x^3-u_Dx^2+\frac{1}{4}\Lambda_D^4x\,,\quad\text{for large $u$}
\end{equation} 
where
\begin{equation}
u_D=\tr\phi_D^2=\frac{2u}{M_{11}}\,,\quad \Lambda_D^4=\frac{16\Lambda_1^4\Lambda_2^4}{M_{11}^2}\,.
\end{equation}
After rescaling $x$ and $y$, (\ref{eqn:largeu}) becomes
\begin{equation}
y^2=x^3-u x^2+\Lambda_1^4\Lambda_2^4x\,,\quad\text{for large $u$}.
\end{equation} 
Then, using the analyticity,  dimensional analysis, and the $\mathbb{Z}_2$ symmetry exchanging $SU(2)_1$ and $SU(2)_2$, we can determine the general form of the $\mathcal{N}=1$ curve,
\begin{equation}\label{eqn:alphacurve}
y^2=x^3-\left(u-\alpha(\Lambda_1^4+\Lambda_2^4)\right)x^2+\Lambda_1^4\Lambda_2^4x\,,
\end{equation} 
up to a numerical constant $\alpha$. We see that the curve becomes singular when either $\Lambda_1=0$ or $\Lambda_2=0$. 

To fix the remaining parameter $\alpha$, we can take a decoupling limit, $\Lambda_2\gg\Lambda_1$. The theory in this limit is approximately an $SU(2)_1$ gauge theory coupled to the adjoint field $\widetilde{\phi}$ and three singlets $\det\Phi_1^2$, $\det\Phi_2^2$, $\tr\Phi_1\Phi_2$. These fields satisfy the following quantum constraint \cite{Seiberg:1994bz}
\begin{equation}\label{eqn:constraint}
u+E^2\widetilde{u}=\Lambda_2^4 \,,
\end{equation}
where $E$ is a dimensionful normalization, and 
\begin{equation}
\widetilde{u}=\tr\widetilde{\phi}^2, \quad \widetilde{\phi}=\frac{1}{E}\left(\Phi_1\Phi_2-\frac{1}{2}\tr\Phi_1\Phi_2\right)\,.
\end{equation}
Decoupling the singlets, the theory is approximately pure $\N=2$ gauge theory with holomorphic scale $\Lambda_1$. The low-energy dynamics is encoded in the curve
\begin{equation}
y^2=x^3-\widetilde{u}x^2+\frac{1}{4}\Lambda_1^4x\,,
\end{equation} 
which becomes singular at $\widetilde{u}=\pm\Lambda_1^2$, or equivalently at $u=\Lambda_2^4\pm E^2 \Lambda_1^2$.
On the other hand, \eqref{eqn:alphacurve} is singular at $u=\alpha\left(\Lambda_1^4+\Lambda_2^4\right)\pm2\Lambda_1^2\Lambda_2^2$. Thus, we find that $\alpha=1$, and therefore the $\mathcal{N}=1$ curve $\mathcal{X}$ of this theory is given by
\begin{equation}\label{eqn:SIcurve}
y^2=x^3-(u-\Lambda^4_1-\Lambda^4_2)x^2+\Lambda^4_1\Lambda^4_2x\,.
\end{equation}
Interestingly, the solution of this $SU(2)\times SU(2)$ $\mathcal{N}=1$ gauge theory is isomorphic to that of the $SU(2)$ $\mathcal{N}=2$ gauge theory, with the isomorphism $\mathfrak{X}\iso\mathfrak{X}_{\N=2}\iso \mathbb{H}/\Gamma_0(4)$ given by the map $f:\left(u_{\N=2},\Lambda_{\N=2}\right)\mapsto \left(u-\Lambda_1^4-\Lambda_2^4,\Lambda_1\Lambda_2\right)$.

\section{IR curves of class $\mathcal{S}_k$ theories}
\label{sec:SI4classSk}

\begin{figure}[t]
\begin{center}
\includegraphics[scale=0.60]{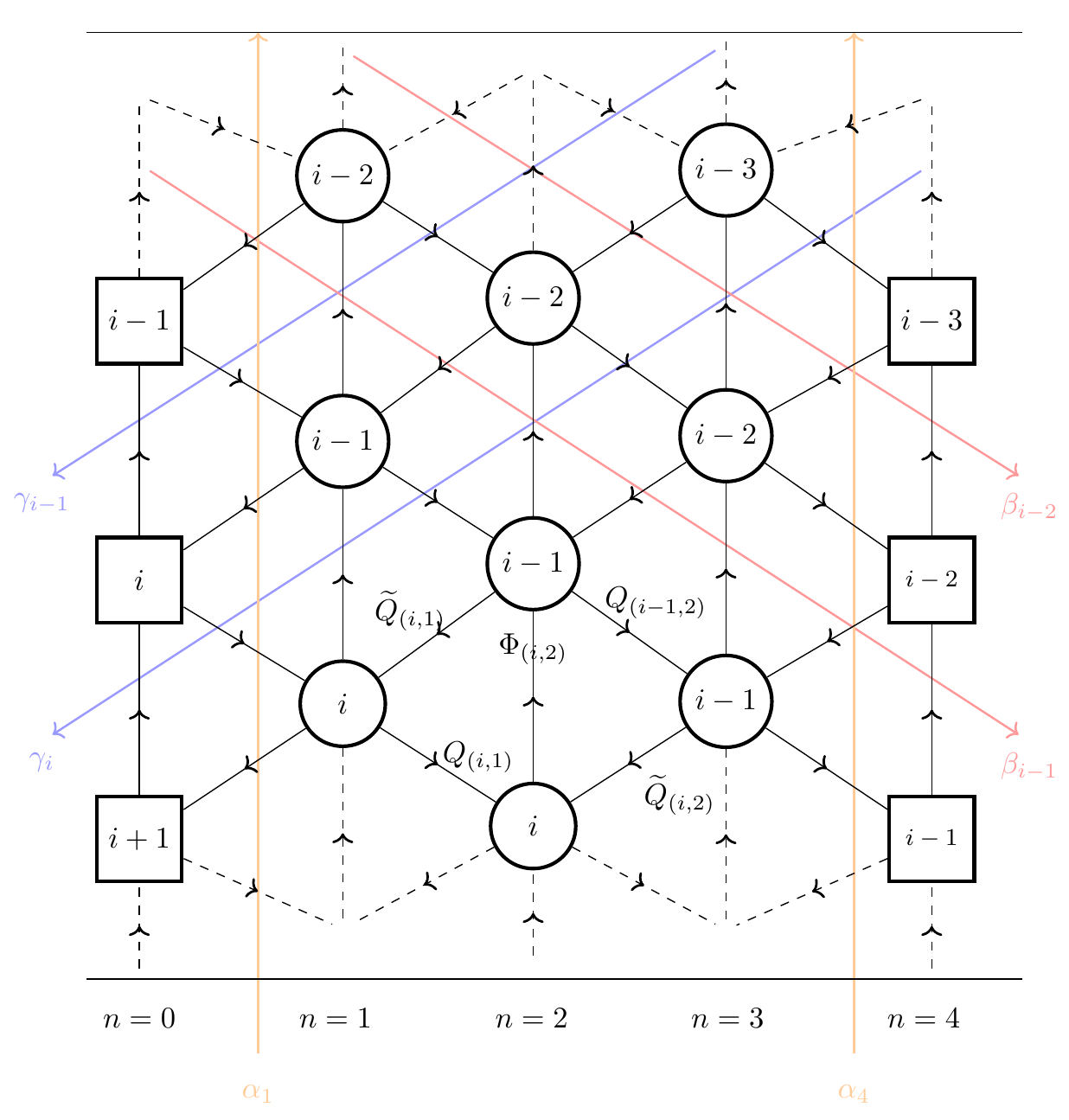}
\end{center}
\caption{\it Quiver for a theory of class $\mathcal{S}_k$. The gauge nodes are labelled by $(i,n)$, where $i=1,...,k$ is the index for the $\mathbb{Z}_k$ orbifold, and $n=1,\cdots,\ell-3$ is the label from the $\mathcal{N}=2$ mother theory. In this example we set $\ell-2=4$. \label{fig:quiverSK}}
\end{figure}

In this section we derive the $\mathcal{N}=1$  curves for the `core theories' of class $\mathcal{S}_k$, namely those associated to spheres with two maximal and $\ell-2$ minimal punctures. 
See Figure \ref{fig:quiverSK} for the $\ell-2=4$ example and notation.

Before plunging into the details, a short review of a few facts and notations about class $\mathcal{S}_k$ is in order.
Theories of class $\mathcal{S}_k$ arise as twisted compactifications  of the 6d $\mathcal{N}=(1,0)$ SCFT which is the worldvolume theory on $N$ coincident M5-branes probing the transverse $A_{k-1}$ singularity.
As discussed in detail in \cite{Bourton:2020rfo} these compactifications enjoy an
$\mathcal{N}=1$ $\mathfrak{u}(1)_r$ R-symmetry, which is given by the linear combination
\begin{equation}
r = \frac{2}{3} \left( 2R_{\mathcal{N}=2}-r_{\mathcal{N}=2} \right) \,,
\end{equation}
where $\{R_{\mathcal{N}=2},r_{\mathcal{N}=2}\}$ are 
the Cartan generators  of the $\mathfrak{su}(2)_R \oplus \mathfrak{u}(1)_r$ $\mathcal{N}=2$ R-symmetry algebra.
Another independent linear combination
\begin{equation}
q_t=R_{\mathcal{N}=2}+r_{\mathcal{N}=2}
\end{equation}
generates the $\mathfrak{u}(1)_t$ symmetry inside the  $\mathfrak{u}(1)_t\oplus\mathfrak{u}(1)^{\oplus k-1}_{\beta}\oplus\mathfrak{u}(1)^{\oplus k-1}_{\gamma}$  `intrinsic' global symmetry carried by all theories of class $\mathcal{S}_k$  \cite{Gaiotto:2015usa}. 
We summarize the field content and the transformation properties in Table \ref{tab:fieldsSK}.
\begin{table}[]
{\renewcommand{\arraystretch}{1.2}
\centering
\resizebox{\textwidth}{!}{
\begin{tabular}{|c||c|c|c|c|c|c|c|c|}
\hline
& $SU(N)_{(i,n-1)}$  & $SU(N)_{(i,n)}$  & $SU(N)_{(i-1,n)}$ & $U(1)_t$ & $U(1)_{\alpha_n}$ & $U(1)_{\beta_{i+1-n}}$ & $U(1)_{\gamma_i}$ \\
\hline
\hline
$V_{(i,n)}$ & $\mathbf{1}$ & Adj & $\mathbf{1}$ & 0 & 0 & 0 & 0\\
\hline
$\Phi_{(i,n)}$ & $\mathbf{1}$ & $\mathbf{N}$ & $\overline{\mathbf{N}}$ & -1 & 0 & -1 & +1\\
\hline
$Q_{(i,n-1)}$ & $\mathbf{N}$ & $\overline{\mathbf{N}}$ & $\mathbf{1}$ & +1/2 & -1 & +1 & 0\\
\hline
$\widetilde{Q}_{(i,n-1)}$ & $\overline{\mathbf{N}}$ & $\mathbf{1}$ & $\mathbf{N}$ & +1/2 & +1 & 0 & -1\\
\hline 
\end{tabular}
}
}
\caption{\it Field content and transformation properties of a class $\mathcal{S}_k$ theory  \label{tab:fieldsSK}}
\end{table}
For further details the reader is encouraged to look at \cite{Bourton:2020rfo}, the notation in which we closely follow.

\subsection{Classical Analysis}
The theory admits a rather intricate phase structure. However, different phases do not mix because they may be differentiated using the $U(1)_t$ symmetry, as we have shown in \cite{Bourton:2020rfo}. Thus, we may restrict our attention to the gemeralized Coulomb branch defined by giving nonzero vevs to $\Phi$'s while $Q,\widetilde{Q}$ have vanishing vevs. Generically the gauge group is spontaneously broken from  $SU(N)^{k\ell}$ down to $U(1)^{(N-1)\ell}$.

In terms of $\Phi_n:=\prod_{i=1}^k\Phi_{(i,n)}$, the Coulomb branch may be parametrized by vevs of the following gauge-invariant operators of dimension $lk$, 
\begin{equation}\label{eqn:uln}
u_{lk,n}:=\begin{cases}\frac{1}{l}\tr\left(\Phi_n-\frac{1}{N}\tr\Phi_n\right)^l& 2\leq l\leq N\\
\tr\Phi_n&l=1
\end{cases}  
\end{equation}
for each $n=1,\cdots,\ell-3$, 
as well as the `baryonic' gauge-invariant operators of dimension $N$,
\begin{equation}\label{eqn:bin}
B_{i,n}:=\frac{1}{(N!)^2}\epsilon_{{a_1}\cdots {a_N}}\epsilon^{{b_{1}}\cdots {b_{N}}}\Phi_{(i,n)b_1}^{{a_1}}\cdots\Phi_{(i,n)b_N}^{{a_N}}=\det\Phi_{(i,n)} 
\end{equation}
for each $i=1,\cdots,k$ and $n=1,\cdots,\ell-3$.
We also introduce 
\begin{equation}
(M_{i+1,n})^a_c:=\Phi_{(i,n)b}^{a}\Phi_{(i+1,n)c}^{b}\,.
\end{equation}
Classically these operators obey the relations 
\begin{equation}\label{eqn:classicalrelation}
\det M_{i+1,n}-B_{i,n}B_{i+1,n}=0\,,
\end{equation}
which are modified quantum mechanically as \cite{Seiberg:1994bz}
\begin{equation}
\det M_{i+1,n}-B_{i,n}B_{i+1,n}=\Lambda^{2N}_{i+1,n} \, .
\end{equation}
Here $\Lambda^{2N}_{i+1,n}$ is the dynamically generated holomorphic scale associated to the $(i+1,n)^{\text{th}}$ gauge group. Its expression in terms of the coupling constants $q_{(j,n)}$ and the mass parameters will be discussed in detail later.

Note that there is an over-parametrization since the $u_{lk,n}$ and $B_{i,n}$ are not all independent but rather related by applying the Cayley-Hamilton theorem to the matrix $\Phi_n$:
\begin{equation}\label{eqn:CHtheorem}
p(\Phi_n)=\sum_{l=1}^Nc_{l,n}\Phi_n^l+(-1)^N\det\Phi_n\mathbb{I}_{N}=0 \, ,
\end{equation} 
where $c_{N,n}=1$ and for $l=1,\cdots,N-1$
\begin{equation}
c_{N-l,n}=\frac{(-1)^l}{l!}\mathbf{B}_l\left(\mathfrak{u}_{k,n},-1!\mathfrak{u}_{2k,n},2!\mathfrak{u}_{3k,n},\cdots,(-1)^{l-1}(l-1)!\mathfrak{u}_{lk,n}\right)\,,
\end{equation}
with $\mathbf{B}_l$ the $l^{\text{th}}$ complete exponential Bell polynomial. Here we have defined $\mathfrak{u}_{lk,n}:=\tr\Phi_n^l$, which are related to $u_{lk,n}$ by
\begin{equation}
u_{lk,n}=\frac{1}{l}\sum_{p=0}^l\binom{l}{p}\left(-\frac{1}{N}\mathfrak{u}_{k,n}\right)^p\mathfrak{u}_{(l-p)k,n}\,.
\end{equation} 
Taking the trace of \eqref{eqn:CHtheorem} implies, for generic $\Phi_{(i,c)}$, a single relation between \eqref{eqn:uln} and \eqref{eqn:bin}
\begin{equation}
\tr p(\Phi_n)=\sum_{l=1}^Nc_{l,n}\mathfrak{u}_{lk,n}+(-1)^NN\prod_{i=1}^kB_{i,n}=0\,.
\end{equation}
In particular, this implies that $u_{Nk,n}$ can be completely written in terms of the $B_{i,n}$ and the $u_{lk,n}$ $1\leq l\leq N-1$.
Hence, the coordinate ring of the Coulomb branch for $k\geq2$ is expected to be a freely generated ring of dimension $(3g-3+\ell)(k+N-1)$,
\begin{equation}
\mathbf{CB}=\mathbb{C}[u_{lk,n},B_{i,n}]\,,\quad \begin{aligned}
&l\in\{1,2,\cdots,N-1\}\,,\\
&i\in\{1,2,\cdots,k\}\,,\\
&n\in\{1,2,\cdots,3g-3+\ell\}\,.
\end{aligned}
\end{equation}

\subsection{Mass Parameters}
We may regard the masses corresponding to the flavour symmetries for \newline$U(N)_{(i,0)}=U(N)_{(i,L)}$ and $U(N)_{(i,\ell+1)}=U(N)_{(i,R)}$ as expectation values for background superfields $\Phi_{(i,0)}=M^L_i$, $\Phi_{(i,\ell+1)}=M^R_i$. Hence we may construct flavour symmetry invariant combinations of masses in the same fashion. We define
\begin{gather}
\mu^{L}_{lk}:=\tr\left(\prod_{i=1}^kM_i^L\right)^l\,,\quad \mu^{R}_{lk}:=\tr\left(\prod_{i=1}^kM_i^R\right)^l\,,\\ J^L_i:=\det M_i^L\,,\quad J^R_i:=\det M_i^R\,,
\end{gather}
for $1\leq l\leq N$. Moreover, $M_{i}^{L/R}$ may be diagonalized by $SU(N)$  transformations such that they take the form 
\begin{equation}
M_{i}^{L/R}=\diag\left(m_{i,1}^{L/R},\dots,m_{i,N}^{L/R}\right)\,.
\end{equation}
Additionally, when $\ell=1$, the $SU(N)^{2k}$ flavour symmetry enhances to $SU(2N)^k$  and in that case we find it convenient to combine the mass parameters as 
\begin{equation}
M_i:=M_i^L\oplus M_i^R :=\diag\left(m_{(i),1},\dots,m_{(i),2N}\right)\,,
\end{equation}
and to write the invariants as 
\begin{equation}\label{eqn:massparam2}
\mu_{lk}:=\tr\left(\prod_{i=1}^kM_i\right)^l\,,\quad J_i:=\det M_i\,,
\end{equation}
where now $1\leq l\leq 2N$.

%After turning on  vevs in the UV each $U(1)$ factor is decoupled and the corresponding coupling matrix $\tau_{ab,n}^{\text{UV}}$ is diagonal with each elements associated to the $SU(N)_{(i,n)}$ gauge couplings by \begin{equation} \tau_{ab,n}^{\text{UV}}=2\delta_{a,b}\sum_{i=1}^k\tau_{(i,n)}\end{equation}where $\tau_{(i,n)}$ are the holomorphic couplings appearing in the Lagrangian. 

\subsection{Curves for \texorpdfstring{$N=k=2$}{N=k=2}}

For the sake of starting smoothly, we will first compute the curve for the simplest example, namely the quiver with $N=k=2$ and $\ell=1$. We will drop the index $n$ for compactness, e.g. here $\Phi_{(i)}:=\Phi_{(i,n)}=\Phi_{(i,1)}$. 
\subsubsection{Diagonal Limit}
Initially let us consider the limit where, say, $\Phi_{(1)}$ gets a large diagonal vev $\left<\Phi_{(1)}\right>=\text{diag}(a,-a)$.
By examining the Lagrangian one sees that the gauge group is broken down to a $SU(2)_D$ diagonal subgroup just as in \cite{Intriligator:1994sm}, under which both bifundamentals decompose into an adjoint and a singlet, among which the adjoint associated to $\Phi_{(1)}$ becomes the longitudinal modes of the massive $(SU(2)_{(1)}\times SU(2)_{(2)})/SU(2)_D$ gauge bosons, while the (anti)fundamentals of the $SU(2)$'s decompose into (anti)fundamentals of the $SU(2)_D$.
The uneaten adjoint gives $\phi=\Phi_{(2)}-\frac{1}{2}\tr\Phi_{(2)}$. Below the scale $|a|$ the quarks $Q_{(1,0)}$, $Q_{(2,1)}$, $\tilde{Q}_{(1,0)}$, $\tilde{Q}_{(1,1)}$ can be integrated out, and the superpotential (without the singlets) is then ``diagonal''\footnote{The ``diagonal'' quark masses can be calculated by solving the F-term's for $Q_{(1,0)}$, $Q_{(2,1)}$, $\tilde{Q}_{(1,0)}$, $\tilde{Q}_{(1,1)}$.}. 
In general there will also be a singlet $u_2=\tr\Phi_{(1)}\Phi_{(2)}$ associated to which there can be a mass deformation to the diagonal superpotential $W_D$,
\begin{equation}
W_D\to W_D+m_su_2 \,.
\end{equation}
Then below $|m_s|$ the singlet $u_2$ can be integrated out and can be neglected when we deal with the low energy dynamics of gauge field \cite{Intriligator:1994sm,Csaki:1997zg,Tachikawa:2011ea}.
Hence the low energy effective theory is $\mathcal{N}=2$ QCD with $N_f=4$ flavors \cite{Leigh:1996ds} whose solution is encoded in the curve \cite{Seiberg:1994rs,Argyres:1995wt}
\begin{equation}
y^2=(x^2-u_D)^2-\frac{4q_D}{(1+q_D)^2}\prod_{j=1}^4\left(x+\tilde{\mu}_j-\frac{q_D}{2(1+q_D)}\sum_{f=1}^4\tilde{\mu}_f\right)\,.
\end{equation}
Here the curve is written in the quartic form. We can express $u_D$, $\tilde{\mu}_j$, and the exponentiated coupling $q_{D}$ for the $SU(2)_D$  in terms of the parameters in the $\mathcal{N}=1$ theory as
\begin{eqnarray} 
u_D&:=&\frac{1}{2}\tr\phi^2=u_4/a^2=(\mathfrak{u}_4-\mathfrak{u}_2^2)/2a^2, \\
\tilde{\mu}_j &=& m_{(1),j}m_{(2),j}/a, \\
q_{D}&=&e^{2\pi i\tau_{D}}=q_{(1)}q_{(2)}.
\end{eqnarray}
After rescaling $x\to x/a$, $y\to y/{a^2}$ and substituting in the above relations we have
\begin{equation}\label{eqn:N2curve}
y^2=(x^2-u_4)^2-\frac{4q}{(1+q)^2}\prod_{j=1}^4\left(x+m_{(1),j}m_{(2),j}-\frac{q}{2(1+q)}\mu_2\right) \,,
\end{equation}
where $q:=q_{(1)}q_{(2)}$ and $\mu_2$ is defined as in \eqref{eqn:massparam2},
\begin{equation}
\mu_2 = \sum_{j=1}^4 m_{(1),j}m_{(2),j} \,.
\end{equation}
Now consider integrating out all of the flavours from this $\mathcal{N}=2$ curve, leaving us with pure $SU(2)$ $\mathcal{N}=2$ gauge theory. As indicated in \cite{Argyres:1995wt}, we should hold fixed the relation
\begin{equation}
\Lambda^4_{D}=\frac{4q_{D}}{(1+q_{D})^2}\prod_{j=1}^4\tilde{\mu}_j=\frac{4q_{(1)}q_{(2)}}{(1+q_{(1)}q_{(2)})^2}\prod_{j=1}^4\frac{m_{(1),j}m_{(2),j}}{a} \,.
\end{equation}
On the other hand, we know from \cite{Intriligator:1994sm}  that \footnote{Note that to save on various factors of $\sqrt{2}$ we define $u_D=\frac{1}{2}\tr\phi^2$ instead of $u_D=\tr\phi^2$, accounting for factor $4$ instead of $16$.}
\begin{equation}
\Lambda_{D}^4=4\frac{\Lambda^4_{(1)}\Lambda^4_{(2)}}{a^4}
\end{equation}
should be held fixed. Equating them implies that
\begin{equation}
\Lambda^4_{(1)}\Lambda^4_{(2)}=\frac{q_{D}}{(1+q_{D})^2}\prod_{j=1}^4m_{(1),j}m_{(2),j}=\frac{q_{(1)}q_{(2)}}{(1+q_{(1)}q_{(2)})^2}J_1J_2
\end{equation}
should be held fixed in the limit, with $J_i$ defined in \eqref{eqn:massparam2}.
Because of the symmetry between two gauge groups, we must have the matching condition
\begin{equation}\label{eqn:matching}
\Lambda^4_{(i)}=\pm \frac{q_{(i)}}{1+q_{(1)}q_{(2)}}J_i.
\end{equation}
Positivity of $\mathrm{Re}\Lambda^4_{(i)}$ demands that we take the plus sign.

Now we can write down the most general ansatz for the curve, which is both polynomial in masses and Coulomb moduli, and is compactible with all of the symmetries and the diagonal limit,
\begin{equation}\label{eqn:gencurve}
\begin{aligned}
y^2=&\left(x^2-u_4+a_{12}J_1+a_{21}J_2+b\mu_2^2+c\mu_2 u_2+d\mu_4\right)^2\\
&-\frac{4q_{(1)}q_{(2)}}{(1+q_{(1)}q_{(2)})^2}\prod_{j=1}^4\left(x+m_{(1),j}m_{(2),j}-\frac{q_{(1)}q_{(2)}}{2(1+q_{(1)}q_{(2)})}\mu_2\right)\,,
\end{aligned}
\end{equation}
where $a_{12}:=a(q_{(1)},q_{(2)})$, $a_{21}:=a(q_{(2)},q_{(1)})$, and $b,c,d$ are all symmetric functions in $q_{(1)},q_{(2)}$. Note however that we may immediately restrict the dependence of $b,c,d$ on $q_{(1)},q_{(2)}$ by demanding agreement with the curves \cite{Gremm:1997sz,Intriligator:1994sm} upon integrating out some of the flavours. To have a well defined limit, $b,c,d$ must be power series in $q_{(1)}q_{(2)}$ with vanishing constant term.

\subsubsection{$q_{(2)}\gg q_{(1)}$ Limit}
Analogous to the treatment in \cite{Intriligator:1994sm},
we now consider a decoupling limit, $q_{(2)}\gg q_{(1)}$. The low energy effective theory in this limit is described by an $SU(2)$ gauge theory with an adjoint field $\tilde{\phi}=\frac{1}{E}(\Phi_1\Phi_2-\frac{1}{2}\tr\Phi_1\Phi_2)$ and three singlets $B_1,B_2$ and $u_2$. Here $E$ is a chosen energy scale. There are also singlets involving fundamental $Q,\widetilde{Q}$'s which do not appear in the curve for reasons discussed earlier. If we are only interested in the effective gauge coupling in the Coulomb phase, we can neglect these singlets and the theory in this limit is approximately an $\mathcal{N}=2$ $SU(2)_1$ $N_f=4$  gauge theory with the exponentiated
coupling $q_{(1)}$ and mass matrix $M_1$.
The fields are constained, implementing the matching relation, by the quantum relation \cite{Seiberg:1994bz}
\begin{equation}\label{eqn:Pfaffian}
u-E^2\tilde{u}=\frac{q_{(2)}}{1+q_{(1)}q_{(2)}}J_2\,,
\end{equation}
with $\tilde{u}=\frac{1}{2}\tr\tilde{\phi}^2$ \,.

To fix $a_{12}$ and $a_{21}$, we need only consider the mass configurations $M_1=0$, $M_2\sim E\mathbb{I}_4$. Then the $\mathcal{N}=2$ theory has an order $4$ singularity, associated to the quarks becoming massless, when $\tilde{u}=0$. For these mass configurations the discriminant of \eqref{eqn:gencurve} has a point of vanishing order $4$ at  
\begin{equation}\label{eqn:singpt}
u_4=\frac{q_{(2)}}{1+q_{(1)}q_{(2)}}J_2\,,
\end{equation}
which implies that
\begin{equation}
a_{21}\equiv \frac{q_{(2)}}{1+q_{(1)}q_{(2)}},\quad a_{12}\equiv \frac{q_{(1)}}{1+q_{(1)}q_{(2)}}\,.
\end{equation} 

The remaining functions $b,c,d$ can be fixed by considering the limit
\begin{equation}
q_{(1)}\to\infty,\quad q_{(2)}\to 0,\quad m_{(2),j}\to\infty\,,
\end{equation}
while holding $q_{(1)}{q_{(2)}}\propto1$ and the matching condition \eqref{eqn:matching} fixed. Then we can integrate out all the massive modes and arrive at the $\mathcal{N}=1$ $SU(2)_1\times SU(2)_2$ quiver theory with $2$ chiral multiplets and $2$ anti-chiral multiplets in $(\mathbf{2},\mathbf{1})$, and a chiral multiplet and an anti-chiral multiplet in $(\mathbf{2},\mathbf{2})$. The resulting curve is well defined only when $b=c=d=0$.\footnote{For example the term $ b(q_{(1)}q_{(2)})\mu_4\to b(1)\infty^2$ explodes unless $b\equiv0$.} 

Hence, the curve is
\begin{align}\label{eqn:curve}
&\begin{aligned}
y^2=&\left(x^2-u_4+\frac{q_{(1)}}{1+q_{(1)}q_{(2)}}J_1+\frac{q_{(2)}}{1+q_{(1)}q_{(2)}}J_2\right)^2\\
&-\frac{4q_{(1)}q_{(2)}}{(1+q_{(1)}q_{(2)})^2}\prod_{j=1}^4\left(x+m_{(1),j}m_{(2),j}-\frac{q_{(1)}q_{(2)}}{2(1+q_{(1)}q_{(2)})}\mu_2\right)\,,
\end{aligned}
\end{align}
or by substituting the relations
\begin{equation}
J_i=\det M_i=\prod_{j=1}^4m_{(i),j}\,,\quad \mu_2=\tr M_1M_2=\sum_{j=1}^4m_{(1),j}m_{(2),j}\,,
\end{equation}
we have
\begin{align}\label{eq:SU(2)l1curve}
y^2=&\left(x^2-u_4+\frac{q_{(1)}}{1+q_{(1)}q_{(2)}} \prod_{j=1}^4m_{(1),j}
+\frac{q_{(2)}}{1+q_{(1)}q_{(2)}}\prod_{j=1}^4m_{(2),j}       \right)^2
\\
&-\frac{4q_{(1)}q_{(2)}}{(1+q_{(1)}q_{(2)})^2}\prod_{j=1}^4\left(x+m_{(1),j}m_{(2),j}-\frac{q_{(1)}q_{(2)}}{2(1+q_{(1)}q_{(2)})}\sum_{j=1}^4m_{(1),j}m_{(2),j}
\right)\,.
\end{align}

\subsubsection{Checks}
It can be immediately verified that our curve \eqref{eqn:curve} reproduces those of \cite{Gremm:1997sz,Intriligator:1994sm}.
It is illustrative to consider the limit $q_{(2)}\gg q_{(1)}$  and check consistency for a few choices of mass deformations:
\paragraph{$M_i=\diag(m,-m,m,-m)$, $M_2\sim E\mathbb{I}_4$}
In this configuration the adjoint field of $SU(2)_1$ is singular at the order $4$ quark singularity $\tilde{u}=m^2$ whilst the corresponding vanishing order $4$ point of \eqref{eqn:curve} is at 
\begin{equation}
\begin{aligned}
u_4&=m^2E^2+\frac{q_{(1)}}{1+q_{(1)}q_{(2)}}m^4+\frac{q_{(2)}}{1+q_{(1)}q_{(2)}}E^4\\
&\approx m^2E^2+\frac{q_{(2)}}{1+q_{(1)}q_{(2)}}E^4\,,
\end{aligned}
\end{equation}
which agrees nicely with \eqref{eqn:Pfaffian}.

\paragraph{$M_1=\diag(m,m,m,0)$, $M_2\sim E\mathbb{I}_4$}
The $\mathcal{N}=2$ curve has an order $3$ quark singularity at $4\widetilde{u}=m^2(2-q_{(1)})^2/(1+q_{(1)})^2= 4m^2+\mathcal{O}(q_{(1)})$. On the other hand \eqref{eqn:curve} has a vanishing order $3$ point at
\begin{equation}
\begin{aligned}
u_4&=\frac{q_{(2)}}{1+q_{(1)}q_{(2)}}E^4+E^2m^2+\frac{3(1-q_{(1)}q_{(2)})}{1+q_{(1)}q_{(2)}}\\
&\approx \frac{q_{(2)}}{1+q_{(1)}q_{(2)}}E^4+4E^2m^2\,,
\end{aligned}
\end{equation}
which is in agreement with \eqref{eqn:Pfaffian}.

\paragraph{$M_1=\diag(m,m,m,m)$, $M_2\sim E\mathbb{I}_4$}
The $\mathcal{N}=2$ curve has an order $4$ quark singularity at $\widetilde{u}=m^2(1-q_{(1)})^2/(1+q_{(1)})^2=m^2+\mathcal{O}(q_{(1)})$. The discriminant of \eqref{eqn:curve} has a zero of degree $4$ located at
\begin{equation}
\begin{aligned}
u_4=&\frac{q_{(1)}}{1+q_{(1)}q_{(2)}}m^4+\frac{q_{(2)}}{1+q_{(1)}q_{(2)}}E^4+\frac{(1-q_{(1)}q_{(2)})^2}{(1+q_{(1)}q_{(2)})^2}m^2E^2\\
\approx& \frac{q_{(2)}}{1+q_{(1)}q_{(2)}}E^4+m^2E^2 \,,
\end{aligned}
\end{equation}
which is again in agreement with \eqref{eqn:Pfaffian}.

\subsection{Curve for \texorpdfstring{$k=2$}{k=2} General \texorpdfstring{$N$}{N}}
The generalization from $N=2$ to general $N$ is rather straightforward. We again consider the diagonal limit and the limit $q_{(2)}\gg q_{(1)}$.

\subsubsection{Diagonal Limit} The same diagonal limit is reached by giving diagonal vev. 
In this limit the theory is again approximately $\mathcal{N}=2$ $SU(N)$ gauge theory with $N_f=2N$ flavours and the adjoint field is $\phi=\Phi_{(2)}-\frac{1}{N}\tr\Phi_{(2)}$. The curve of that theory is given by \cite{Argyres:1995wt}
\begin{equation}
\begin{aligned}
y^2=&\left(x^N-\sum_{l=2}^Nu_{D,l}x^{N-l}\right)^2\\
&-\frac{4q_D}{(1+q_D)^2}\prod_{j=1}^{2N}\left(x+\tilde{\mu}_j-\frac{q_D}{N(1+q_D)}\sum_{m=1}^{2N}\tilde{\mu}_m\right)\,,
\end{aligned}
\end{equation}
where $\tilde{\mu}_j=m_{(1),j}m_{(2),j}/a$, $u_{D,l}:=\frac{1}{l}\tr\phi^l=u_{2l}/a^l$, and $q_{D}=q_{(1)}q_{(2)}$ is associated to the coupling for   $SU(2)_D$. 
After rescaling $x\to x/a$, $y\to y/{a^{2N}}$, and substituting in the above relations we have
\begin{equation}
\begin{aligned}
y^2=&\left(x^{N}-\sum_{l=2}^Nu_{2l}x^{N-l}\right)^2\\
&-\frac{4q_{(1)}q_{(2)}}{(1+q_{(1)}q_{(2)})^2}\prod_{j=1}^{2N}\left(x+m_{(1),j}m_{(2),j}-\frac{q_{(1)}q_{(2)}}{N(1+q_{(1)}q_{(2)})}\mu_2\right)\,.
\end{aligned}
\end{equation}
Now consider integrating out all of the flavours from this $\mathcal{N}=2$ curve to obtain pure $SU(2)$ $\mathcal{N}=2$ gauge theory, holding fixed the relation
 \cite{Argyres:1995wt} 
\begin{equation}
\Lambda^{2N}_{D}=\frac{4q_{D}}{(1+q_{D})^2}\prod_{j=1}^{2N}\tilde{\mu}_j=\frac{4q_{(1)}q_{(2)}}{(1+q_{(1)}q_{(2)})^2}\prod_{j=1}^{2N}\frac{m_{(1),j}m_{(2),j}}{a} \,.
\end{equation}
Meanwhile, we should also hold fixed  \cite{Csaki:1997zg}
\begin{equation}
\Lambda_{D}^{2N}=4\frac{\Lambda^{2N}_{(1)}\Lambda^{2N}_{(2)}}{a^{2N}}\,.
\end{equation}
Equating them implies that
\begin{equation}
\Lambda^{2N}_{(1)}\Lambda^{2N}_{(2)}=\frac{q_{(1)}q_{(2)}}{(1+q_{(1)}q_{(2)})^2}J_1J_2
\end{equation}
should be held fixed under the limit, with $J_i$ defined in \eqref{eqn:massparam2}.
Due to the symmetry between two gauge groups, we must have that
\begin{equation}
\Lambda^{2N}_{(i)}=\pm \frac{q_{(i)}}{1+q_{(1)}q_{(2)}}J_i\,,
\end{equation}
and we should pick the plus sign if we want to have $\mathrm{Re}\Lambda^{2N}_{(i)}>0$.

The most general ansatz for the curve takes the form
\begin{equation}\label{eqn:gencurveN}
\begin{aligned}
y^2=&\left(x^N-\sum_{l=2}^N\left[u_{2l}+f_l\left(\mathfrak{u}_n,\mu_{m};q_{(1)}q_{(2)}\right)\right]x^{N-l}+a_{12}J_1+a_{21}J_2\right)^2\\
&-\frac{4q_{(1)}q_{(2)}}{(1+q_{(1)}q_{(2)})^2}\prod_{j=1}^{2N}\left(x+m_{(1),j}m_{(2),j}-\frac{q_{(1)}q_{(2)}}{N(1+q_{(1)}q_{(2)})}\mu_2\right),
\end{aligned}
\end{equation}
where $f_l\left(u_{2n},\mu_{m};q_{(1)}q_{(2)}\right)$ is a function with mass dimension (or equivalently R-charge) $2l$ and satisfies $\lim_{\mu_m\to0}f_l\left(u_{2n},\mu_{m};q_{(1)}q_{(2)}\right)=0$. In other words $f_{l}$ is always subdominant compared to $u_{2l}$ in the large $u_{2l}$ limit. Following the same argument as before, we know that the dependence of $f_l$ on the holomorphic couplings can only be of the combination $q_{(1)}q_{(2)}$.

\subsubsection{$q_{(2)}\gg q_{(1)}$ Limit}
To determine the remaining parameters in \eqref{eqn:gencurveN}, we again consider the limit $q_{(2)}\gg q_{(1)}$. In that limit, the theory is approximately $\mathcal{N}=2$ $SU(N)_1$ gauge theory with $N_f=2N$ fundamental hypermultiplets and the exponentiated
coupling $q_{(1)}$. There is an adjoint field $\widetilde{\phi}=\frac{1}{E}\left(\Phi_{(1)}\Phi_{(2)}-\frac{1}{N}\tr\Phi_{(1)}\Phi_{(2)}\right)$.  There is also a quantum modified constraint on moduli space \cite{Seiberg:1994bz}
\begin{equation}\label{eqn:SeibergCons}
\det\Phi_{(1)}\Phi_{(2)}-B_1B_2=\frac{q_{(2)}}{1+q_{(1)}q_{(2)}}J_2\,.
\end{equation} 
This is implemented on the curves by
\begin{equation}\label{eqn:quantcon}
\widetilde{u}_l=\begin{cases}\frac{u_{2l}}{E^l}&l\neq N\\
\frac{1}{E^l}\left(u_{2N}+\frac{q_{(2)}}{1+q_{(1)}q_{(2)}}J_2\right)&l= N
\end{cases}\,.
\end{equation}
Now we may fix $a_{12}(q_{(1)},q_{(2)})=a_{21}(q_{(2)},q_{(1)})$. The massless $M_1=0$ $\mathcal{N}=2$, $N_f=2N$ theory is singular when $\widetilde{u}_l=0$. On the other hand, our curve \eqref{eqn:gencurveN} with $M_1=0$ is singular when 
\begin{equation}
u_{2l}=\begin{cases}0&l\neq N\\
a_{21}J_2&l= N
\end{cases}\,.
\end{equation}
Comparison with \eqref{eqn:quantcon} implies $a_{21}=\frac{q_{(2)}}{1+q_{(1)}q_{(2)}}$. We may then follow the argument that we used below equation \eqref{eqn:singpt} to immediately set $f_l\equiv0$ for all $l$. Hence the curve  is
\begin{equation}\label{eqn:curveN}
\begin{aligned}
y^2=&\left(x^N-\sum_{l=2}^Nu_{2l}x^{N-l}+\frac{q_{(1)}}{1+q_{(1)}q_{(2)}}J_1+\frac{q_{(2)}}{1+q_{(1)}q_{(2)}}J_2\right)^2\\
&-\frac{4q_{(1)}q_{(2)}}{(1+q_{(1)}q_{(2)})^2}\prod_{j=1}^{2N}\left(x+m_{(1),j}m_{(2),j}-\frac{q_{(1)}q_{(2)}}{N(1+q_{(1)}q_{(2)})}\mu_2\right)\,.
\end{aligned}
\end{equation}

\subsection{Curve for general \texorpdfstring{$N$}{N} \& \texorpdfstring{$k$}{k}}
The generalization to arbitrary $k$ is largely the same. It is the extension of \cite{Csaki:1997zg} to include flavours. The only new phenomena is the implementation of the quantum relation \eqref{eqn:SeibergCons} for the quiver. We will simply state the result. The curve may be written as
\begin{equation}\label{NK}
\begin{aligned}
y^2=&\left(\sum_{l=1}^Nc_{l}\mathfrak{u}_{lk}x^{N-l}+(-1)^N\prod_{i=1}^kB_{i}+\left[B_iB_{i+1}\to\frac{q_{i+1}}{1+q}J_{i+1}\right]\right)^2\\
&-\frac{4q}{(1+q)^2}\prod_{j=1}^{2N}\left(x+m_j-\frac{q}{N(1+q)}\mu_k\right)\,,
\end{aligned}
\end{equation}
where the $c_l$ are defined by \eqref{eqn:CHtheorem}, $q:=\prod_{i=1}^kq_{(i)}$, $m_j:=\prod_{i=1}^km_{(i),j}$. Finally the brackets $[\cdot]$ mean to replace the pairs $B_iB_{i+1}$ appearing in $(-1)^N\prod_{i=1}^kB_{i}$ with the corresponding mass condition in all possible ways. For example at $k=4$ the brackets should be read as
\begin{equation}
\begin{aligned}
(-1)^N\left(B_iB_{i+1}\to\Lambda^{2N}_{i+1}\right)=&\Lambda^{2N}_2B_3B_4+B_1\Lambda^{2N}_3B_4+B_1B_2\Lambda^{2N}_4\\
&+\Lambda^{2N}_1B_2B_3+\Lambda^{2N}_2\Lambda^{2N}_4+\Lambda^{2N}_1\Lambda^{2N}_3\,,
\end{aligned}
\end{equation}
where we used $\Lambda_i^{2N}=q_{i}J_{i}/(1+q)$.

\section{UV curves of class $\mathcal{S}_k$ theories}
\label{sec:UVcurves}

In this section we will
 derive the UV curves by rewriting our IR curves following the procedure introduced by Gaiotto \cite{Gaiotto:2009we}. 
 Then we will compare our results with the results obtained in \cite{Coman:2015bqq} which are valid only at the orbifold point of the conformal manifold.

\subsection{Review of UV curves of class $\mathcal{S}$ theories}
Let us begin by reviewing the manipulation needed for class $\mathcal{S}$ theories.
The Seiberg-Witten curve of $\N=2$ $SU(N)$ gauge theory with $N_f=2N$ flavors reads  \cite{Argyres:1995wt}
\begin{equation}
y^2=\left(x^N-\sum_{l=2}^Nu_{l}x^{N-l}\right)^2-\frac{4q}{(1+q)^2}\prod_{j=1}^{2N}\left(x-\mathfrak{m}_j\right)\,,
\end{equation}
with $\mathfrak{m}_j$ given in terms of the masses $m_j$ as
\begin{equation}
\mathfrak{m}_j=-m_j+\frac{q}{N(1+q)}\sum_{f=1}^{2N}m_f\,.
\end{equation}
The dimensions of $(x,y)$, which are minus the $U(1)_{r_{\mathcal{N}=2}}$ charges, are $(1,N)$.
After making a change of variables
\begin{equation}
y=-\frac{2t}{1+q}\prod_{j=1}^N\left(x-\mathfrak{m}_j\right)+\left(x^N-\sum_{l=2}^Nu_{l}x^{N-l}\right)\,,
\end{equation}
we obtain
\begin{equation}
\prod_{i=1}^N(x-\mathfrak{m}^L_i)t^2-(1+q)\left(x^N-\sum_{l=2}^Nu_lx^{N-l}\right)t+q\prod_{i=1}^{N}(x-\mathfrak{m}^R_i)=0 \,,
\end{equation}
which is the natural expression obtained by lifting the Type IIA string theory construction using the D4/NS5 brane system to M-theory \cite{Witten:1997sc}.
If we write $x=tz$, this curve is brought to the canonical form introduced by Gaiotto \cite{Gaiotto:2009we},
\begin{equation}
z^N+\sum_{i=1}^Nz^{N-l}\phi_l(t)=0\,,
\end{equation}
where $\phi_l(t)dt^l$ are degree $l$ differentials on the Riemann surface $\mathcal{C}\xleftarrow{1:N}\mathcal{X}$, $t$ is a local coordinate on $\mathcal{C}$ and $(z,t)$ on $T^*\mathcal{C}$. In the massless case, $m_i=0$, the curve is
\begin{equation}
z^N+\sum_{l=2}^Nz^{N-l}\frac{(1+q)u_l}{t^{l-1}(t-1)(t-q)}=0\,.
\end{equation}
In the massive case, we have
\begin{equation}
z^N+\sum_{l=1}^Nz^{N-l}\frac{t^{2}f_l(\mathfrak{m}^{L})+t(1+q)u_l+qf_l(\mathfrak{m}^R)}{t^l(t-1)(t-q)}=0\,,
\end{equation}
where $u_1=0$ and $f_l(\mathfrak{m}^{L/R})$ is given by the expansion
\begin{equation}\label{flm}
\prod_{i=1}^N(x-\mathfrak{m}^{L/R}_i)=x^N+\sum_{l=1}^Nx^{N-l}f_l(\mathfrak{m}^{L/R})\,.
\end{equation}
Accordingly, the differentials are
\begin{equation}
\phi_l(t)dt^l=\frac{t^{2}f_l(\mathfrak{m}^{L})+t(1+q)u_l+qf_l(\mathfrak{m}^R)}{t^l(t-1)(t-q)}dt^l\,.
\end{equation}
Notice that at $t=0$ and $t=\infty$, $\phi_l$ has poles of order $l$.
 These are interpreted as maximal punctures. On the other hand, $\phi_l$ has simple poles at $t=1,q$, and these are the locations of the minimal punctures. This is in complete agreement with the findings of  \cite{Coman:2015bqq}.

\subsection{Class $\mathcal{S}_k$}
Let us extend the above manipulations to the theories of class $\mathcal{S}_k$. We first consider the case when $k=2$. Recall that the curve \eqref{eqn:curveN} is given by
\begin{equation}
\begin{aligned}
y^2=&\left(x^N-\sum_{l=2}^Nu_{2l}x^{N-l}+\frac{q_{(1)}}{1+q_{(1)}q_{(2)}}J_1+\frac{q_{(2)}}{1+q_{(1)}q_{(2)}}J_2\right)^2\\
&-\frac{4q_{(1)}q_{(2)}}{(1+q_{(1)}q_{(2)})^2}\prod_{j=1}^{2N}\left(x+m_{(1),j}m_{(2),j}-\frac{q_{(1)}q_{(2)}}{N(1+q_{(1)}q_{(2)})}\mu_2\right)\,.
\end{aligned}
\end{equation}
The dimensions of $(x,y)$ are $(2,2N)$. If we perform the change of variables
\begin{equation}
y=-\frac{2t}{1+q}\prod_{j=1}^N\left(x-\mathfrak{m}_j\right)+\left(x^N-\sum_{l=2}^Nu_{2l}x^{N-l}+\sum_{i=1}^2\frac{q_{(i)}}{1+q}J_i\right)\,,
\end{equation}
with
\begin{equation}
q=q_{(1)}q_{(2)},\quad \mathfrak{m}_j:=-m_{(1),j}m_{(2),j}+\frac{q}{N(1+q)}\mu_2\,,
\end{equation}
the curve becomes
\begin{equation}
\prod_{i=1}^N(x-\mathfrak{m}^L_i)t^2-\left(1+q\right)\Big(x^N-\sum_{l=2}^Nu_{2l}x^{N-l} +\sum_{i=1}^2\frac{q_{(i)}}{1+q}J_i\Big)t\\
+q\prod_{i=1}^{N}(x-\mathfrak{m}^R_i) = 0 \,.
\end{equation}
Following the analysis of class $\mathcal{S}$ theories, we can bring the curve to the canonical form by writing $x=tz^2$,
\begin{equation}
z^{2N}+\sum_{l=1}^Nz^{2(N-l)}\phi_{2l}(t)=0\,,
\end{equation}
where
\begin{equation}
\phi_{2l}(t)=\begin{cases}
\frac{t^{2}f_1(\mathfrak{m}^{L})+qf_1(\mathfrak{m}^R)}{t(t-1)(t-q)}& l=1\,,\\
\frac{t^{2}f_l(\mathfrak{m}^{L})+t(1+q)u_{2l}+qf_l(\mathfrak{m}^R)}{t^l(t-1)(t-q)}&1<l<N\,,\\
-\frac{q_{(1)}J_1+q_{(2)}J_2}{t^{N-1}(t-1)(t-q)}&l=N\,,
\end{cases}
\end{equation}
and $f_l$ is given by \eqref{flm}.
We see that $(z,t)$ have dimensions $(1,0)$. We can interpret $(z,t)$ as local canonical coordinates in the cotangent bundle  $T^*\mathcal{C}$ of a punctured Riemann surface $\mathcal{C}$.

It is then straightforward to generalize our discussion to arbitrary $k$. The IR curve is given by \eqref{NK}, with the dimensions of $(x,y)$ being $(k,Nk)$. We can again obtain the canonical form of the curve from (\ref{NK}) by first changing of variables from $(x,y)$ to $(x,t)$ and then write $x=tz^k$. The resulting curve takes the form
\begin{equation}
\label{eq:UVgeneralform}
z^{kN}+\sum_{l=1}^Nz^{kN-kl}\phi_{kl}(t)=0\,,
\end{equation}
where
\begin{equation}
\label{eq:meromorphicdiferentials}
\phi_{kl}(t)=\begin{cases}
\frac{t^{2}f_1(\mathfrak{m}^{L})+qf_1(\mathfrak{m}^R)}{t(t-1)(t-q)}& l=1\,,\\
\frac{t^{2}f_l(\mathfrak{m}^{L})+t(1+q)u_{kl}+qf_l(\mathfrak{m}^R)}{t^l(t-1)(t-q)}&1<l<N\,,\\
-\frac{q_{(1)}J_1+q_{(2)}J_2}{t^{N-1}(t-1)(t-q)}&l=N\,.
\end{cases}
\end{equation}
Here $(z,t)$ are again local canonical coordinates in the cotangent bundle  $T^*\mathcal{C}$ of a punctured Riemann surface $\mathcal{C}$, and 
\begin{equation}
q:=\prod_{i=1}^kq_{(i)} \, .
\end{equation}
Notice that, as discussed in \cite{Coman:2015bqq}, $\phi_{kl}$ has order $l$ poles at $t=0,\infty$  and simple poles at $t=0,q$. These signify the locations of the maximal and minimal punctures, respectively. 
The curve simplifies in the absence of mass deformations, and becomes
\begin{equation}
z^{kN}+\sum_{l=2}^N z^{k(N-l)}\frac{(1+q)u_{kl}}{t^{l-1}(t-1)(t-q)}=0\,.
\end{equation}

At the orbifold points of the conformal manifold, $q_{(1)}=\cdots=q_{(k)}$, \eqref{eq:UVgeneralform} and \eqref{eq:meromorphicdiferentials} become those found in \cite{Coman:2015bqq}. \footnote{Notice that the definition of $u_{kl}$ in this paper differs from that in \cite{Coman:2015bqq} by $1+q$ and $f_{l}(\mathfrak{m}^{L/R}) = (-1)^{l} \mathfrak{c}^{(l,k)}_{L/R}$.}

\section{Conclusions}
In this paper we apply the techniques of  Intriligator and Seiberg \cite{Intriligator:1994sm} to derive the IR curves encoding the low energy dynamics of $\mathcal{N}=1$ theories of class $\mathcal{S}_k$  on their generalized Coulomb branch. We then bring these curves to the Gaiotto form, obtaining the UV curve $\mathcal{C}$. We interpret $\mathcal{C}$ as a punctured Riemann surface embedded in $T^*\mathcal{C}$, and the class $\mathcal{S}_k$ theory arises from the compactification of the 6D $(1,0)_{A_{k-1}}$ SCFT on $\mathcal{C}$.
The final forms \eqref{eq:UVgeneralform} and \eqref{eq:meromorphicdiferentials} can be directly compared to the expressions derived in  \cite{Coman:2015bqq}, which is from M-theory and is valid only at the orbifold point of the conformal manifold. We also analyze the meromorphic differentials. 

In the case of the a four-punctured sphere, the positions of the poles of the meromorphic differentials are identical ($t = 0,1,q,\infty$) to the ones in \cite{Coman:2015bqq}, but now with $q=\prod_{i=1}^kq_{(i)}$, being interpreted as the ``average coupling''.
As we move away from the orbifold point,  the novelty is that the mass parameters change as functions of the marginal couplings $q_{(i)}$, in a way we can compute.

In this paper we concentrated on  four-punctured spheres with two maximal and two simple punctures. The curves of other theories with a Lagrangian description of class $\mathcal{S}_k$ should be easy to obtain as well as non-Lagrangian theories which are obtained as strong coupling limit (pants decompositions) of Lagrangian ones. What is more, it would be very interesting to study what are the possible punctures (classification) in class $\mathcal{S}_k$.

The curves, while not encoding enough information to `solve' the low energy effective theory due to the K\"ahler part of the action being unconstrained by holomorphicity,  still encode a large amount of information regarding the theory. In particular, they may prove invaluable for deriving new theories and dualities, as was performed for class $\mathcal{S}$ in \cite{Gaiotto:2009we}. We even believe that through their M-theory interpretation they may even allow us to compute the low energy BPS spectrum.

It is well known that the techniques of instanton counting provide a purely field theoretical derivation of Seiberg-Witten curves of 4D $\mathcal{N}=2$ supersymmetric gauge theories \cite{Nekrasov:2002qd,Nekrasov:2003rj,Nekrasov:2012xe,Zhang:2019msw}. In a separate paper, we will extend this approach to $\mathcal{N}=1$ theories realized by brane box models \cite{Hanany:1997tb,Hanany:1998it}. These theories can be formulated in the $\mathcal{N}=1$ version of the $\Omega$-background, and the partition function on the generalized Coulomb branch can be exactly computed. We can then determine the curves from the partition function in the flat space limit.  This is work that will appear in \cite{Pomoni-Yan-Zhang}.

Finally, very interestingly, our work can be used to study fractons \cite{Haah:2011drr}, emergent topological quasiparticles, through the recent relation discovered in \cite{Razamat:2021jkx}. See also \cite{Geng:2021cmq,Vijay:2015mka,Nandkishore:2018sel,Pretko:2020cko,Seiberg:2020bhn}. As discovered in  \cite{Razamat:2021jkx}  the fracton excitations  live on $\mathcal{C}$ which we are now able to compute.

\acknowledgments

We wish to thank Ioana Coman, Shlomo Razamat and Futoshi Yagi for valuable correspondence.
This research was funded in part by the GIF Research Grant I-1515-303./2019.

\end{document}